\documentclass[sigconf]{acmart}
\AtBeginDocument{%
  }

\usepackage{csquotes}
\usepackage{booktabs,longtable,float, tabularx}
\usepackage{array}
\usepackage{subcaption}
\usepackage{enumitem}

\newcolumntype{P}[1]{>{\raggedright\arraybackslash}p{#1}}

\setcopyright{none}                                 
\renewcommand\footnotetextcopyrightpermission[1]{}  
\begin{document}
\title{(Mis-)Informed Consent: Predatory Apps and the Exploitation of Populations with Limited Literacy}
\author{Muhammad Muneeb Pervez}
\authornote{These authors contributed equally to this work.}
\email{27100022@lums.edu.pk}
\affiliation{%
  \institution{Lahore University of Management Sciences}
  \city{Lahore}
  \state{Punjab}
  \country{Pakistan}
}

\author{Muhammad Qasim Atiq Ullah}
\authornotemark[1]
\email{25100263@lums.edu.pk}
\affiliation{
  \institution{Lahore University of Management Sciences}
  \city{Lahore}
  \state{Punjab}
  \country{Pakistan}
}

\author{Ibrahim Ahmed Khan}
\authornotemark[1]
\email{25100112@lums.edu.pk}
\affiliation{%
  \institution{Lahore University of Management Sciences}
  \city{Lahore}
  \state{Punjab}
  \country{Pakistan}
}

\author{Roshnik Rahat}
\email{26100360@lums.edu.pk}
\affiliation{%
  \institution{Lahore University of Management Sciences}
  \city{Lahore}
  \state{Punjab}
  \country{Pakistan}
}

\author{Aayan Faraz sheikh}
\email{}
\affiliation{%
  \institution{Aitchison college Lahore}
  \city{Lahore}
  \state{Punjab}
  \country{Pakistan}
}

\author{Muhammad Fareed Zaffar}
\email{fareed.zaffar@lums.edu.pk}
\affiliation{%
  \institution{Lahore University of Management Sciences}
  \city{Lahore}
  \state{Punjab}
  \country{Pakistan}
}

\author{Rashid Tahir}
\email{r.tahir@upm.edu.sa}
\affiliation{%
  \institution{University of Prince Murgin}
  \city{Madinah}
  \country{Kingdom of Saudi Arabia}
}

\author{Talal Rahwan}
\email{talal.rahwan@nyu.edu}
\affiliation{%
  \institution{New York University Abu Dhabi}
  \city{Abu Dhabi}
  \country{United Arab Emirates}}

\author{Yasir Zaki}
\email{yasir.zaki@nyu.edu}
\affiliation{%
  \institution{New York University Abu Dhabi}
  \city{Abu Dhabi}
  \country{United Arab Emirates}}

\renewcommand{\shortauthors}{Muhammad Muneeb Pervez et al.}

\begin{abstract}
  Among populations with limited literacy in emerging digital markets, the adoption of mobile phones, combined with comprehension barriers and poor cybersecurity hygiene, has created hidden privacy risks. This paper examines how informed consent is often abused by predatory financial applications, leading to financial scams that disproportionately affect users with low literacy. We focus on predatory loan, gambling, and trading apps, analyzing a dataset of 50 Google Play Store apps to measure how many omit or obfuscate critical privacy disclosures. We also evaluate comprehension gaps among users with low literacy via a targeted user study and assess whether Large Language Model (LLM)-generated summaries, translations, and visual cues can improve consent clarity. Our findings show that 85\% of study participants did not understand basic app permissions, underscoring the urgent need for stronger regulatory oversight and scalable LLM-driven privacy-literacy tools.
\end{abstract}



\maketitle
\vspace{-15pt}
\section{Introduction} 
By 2029, the global smartphone user base is projected to reach 6.1 billion people~\cite{statista2024smartphone}, driving unprecedented adoption among populations with limited literacy. As of 2025, an estimated 754 million adults worldwide are illiterate, two-thirds of whom are women~\cite{unesco_literacy}. Users with low literacy often rely on icon recognition, voice commands, assistance from younger relatives, and memorized tap patterns to navigate their smartphones~\cite{tuli2021actionable}. Although these strategies allow them to use their devices effectively, they can undermine informed consent and have contributed to increasing financial exploitation via predatory financial applications.

Informed consent is a foundational principle in data privacy, formally defined in frameworks such as the General Data Protection Regulation (GDPR) as any ``freely given, specific, informed, and unambiguous indication of the data subject's wishes''~\cite{EuropeanCommission2024_ConsentRequest}. In the context of HCI and mobile permissions, empirical studies emphasize that valid consent requires not merely the mechanical act of clicking ``accept,'' but a demonstrable understanding of the scope and purpose of data collection, a standard often undermined by deceptive interface designs \cite{Santos2020_CookieBanners}. For this study, we operationalize informed consent as ``\textit{the user's ability to meaningfully understand what data an application collects, how it is processed, and the specific risks involved prior to granting access.}'' Moreover, we define users with low/limited literacy as \textit{``adults whose literacy level falls at or below a Level 2 on UNESCO's Literacy Assessment and Monitoring Programme (LAMP). They may have developed coping strategies to manage everyday literacy demands, but their level of proficiency makes it difficult to handle novel tasks such as interpreting multi-sentence passages or abstract concepts. In the mobile context, this often manifests as difficulty decoding legalistic privacy policies, interpreting text-heavy dialog boxes, or distinguishing nuanced permission prompts without external assistance.''} Furthermore, in this paper, we define predatory financial applications (PFAs) as \textit{``mobile apps that leverage misleading or opaque user interfaces, aggressive lending practices (e.g., exorbitant interest rates, hidden rollover fees, or punitive late-payment penalties), or obfuscated permission requests to extract money or personal data from users, often targeting those with limited financial power and digital literacy.''} Finally, we define over-permissioning as \textit{``the application's request for access to device permissions that are not strictly required for the app's advertised core functionality as described in its Play Store listing and user interface.''}

We narrow our focus to loan, gambling, and trading apps, which are often unregulated~\cite{odinet2020predatory, hill2007consumer} and are known for violating user privacy through the misuse of Sensitive Personally Identifiable Information (SPII). These apps are also associated with aggressive and coercive debt collection practices, often resulting in harassment, intimidation, and even blackmail \cite{munyendo2022desperate}. Individuals with limited literacy, who often lack the credit histories required by formal banks, frequently install these apps to secure rapid loans without reviewing the terms and conditions or fully understanding the permissions they grant, which often causes them to become trapped in recurring debt cycles~\cite{carr2001predatory}.

Exploitative gambling applications represent another arm of this predatory ecosystem, targeting compulsive behaviors through tactics like ``free-to-play'' schemes and deceptive starting bonuses. These apps often collect and misuse personal data for fraudulent or non-transparent purposes \cite{Raneri2022}. Similarly, trading apps represent another facet of this predatory profiteering landscape and provide seemingly effortless entry into volatile markets, such as cryptocurrency, often without adequate risk disclosure \cite{acharya2024conning, bartoletti2021cryptocurrency, cui2022privacyglasses}.

The problem is further exacerbated by the Ask On First Use (AOFU) permission model adopted by many mobile applications. Upon installation, apps typically request access to sensitive permissions such as location, contacts, camera, and microphone, requiring users to review and approve or deny each request. However, users with low literacy often grant these permissions without fully understanding what data is being collected and for what purpose.

To establish a baseline of permission awareness, we carried out a survey of blue-collar factory workers with low literacy in Lahore, Punjab, Pakistan, one of the country's largest industrial hubs, where a significant proportion of the population migrates from rural areas for factory employment. In this setting, mobile phone ownership is widespread, but digital and financial literacy remain limited, and many users rely on verbal instructions, social networks, or community intermediaries to navigate technology. These socioeconomic and cultural factors make the region a valuable case study for examining how users with low literacy understand mobile permissions and privacy disclosures in high-risk financial applications. To understand how informed consent is undermined in this context, we conducted a static analysis of 50 predatory financial applications to examine the permissions they request. In the second half of the paper, we present a series of LLM-based interventions designed to simplify privacy information and evaluate their effectiveness in controlled settings. Together, these components allow us to answer the following research questions: \textbf{RQ1:} What is the awareness level and understanding of mobile app permissions and privacy policies among low-literacy users? \textbf{RQ2:} To what extent do predatory loan, gambling, and trading apps request permissions beyond what is necessary? and \textbf{RQ3:} To what extent do LLM-generated summaries, with and without visual cues, enhance comprehension of privacy policies and app permissions among users with limited literacy, particularly in contexts with potential for financial exploitation?

\section {Problem Overview and Related Works}
Understanding how users with low literacy interact with mobile apps requires unpacking both the usability barriers they face and the exploitative dynamics present in certain app categories. Prior research examined various socio-technical challenges in this space, ranging from basic navigation difficulties to issues of informed consent, data privacy, and financial harm. Here, we review key themes in the existing literature that contextualize the problem space and motivate our investigation.
\vspace{-10pt}
\subsection{Informed Consent and Mental Models}
Informed consent in digital privacy has long been criticized as ineffective in practice. Jensen and Potts demonstrated that such policies function poorly as decision-making tools, often serving legal compliance rather than user understanding~\cite{jensen2004privacy}. This disconnect is exacerbated by the ``privacy paradox,'' where users express concern for data protection but fail to act on it due to poor interface design. Shklovski et al. characterized the user experience of mobile data collection as ``creepy'' and ``leaky,'' noting that users often feel a violation of boundaries when apps access data in ways that contradict their mental models~\cite{shklovski2014leakiness}.
Research into these mental models reveals significant gaps between user expectations and system reality. Lin et al. found that users frequently struggle to identify the purpose of permission requests, often assuming functionality that does not exist~\cite{lin2012expectation}. Lin Kyi et al. highlighted that vague purpose strings, such as ``for improving user experience,'' fail to satisfy users' desire for clarity, leading to uninformed acceptance \cite{linkyi2024doesnt}. Even when users exercise control, Habib and Cranor found that existing privacy choice mechanisms are often unusable or buried in submenus~\cite{habib2022evaluating}.

To address these failures, Schaub et al. proposed a design space for effective privacy notices, emphasizing that timing and channel are critical for capturing attention~\cite{schaub2015design}. Harbach et al. further demonstrated that risk communication is significantly improved when abstract data points are translated into personal, concrete examples (e.g., showing the specific photos an app can access)~\cite{harbach2014using} which directly motivated this paper's visual intervention strategies.

While prior research demonstrates that even high-literacy users frequently fail to comprehend permissions, often due to `privacy fatigue,' habituation, or legal text complexity, the barriers faced by users with limited literacy are structural rather than behavioral. Unlike high-literacy users who may choose to ignore privacy disclosures, the population in this study frequently cannot decode them due to text-heavy interface designs. This vulnerability is uniquely compounded by a reliance on `proxy installers' (e.g., family members), a practice that often removes the primary user from the consent loop entirely during the critical installation phase, which is a dynamic rarely observed in studies of digitally fluent populations.

\subsection{Users with Low Literacy and Digital Access}
Users from low-literacy demographics often lack nuanced understanding of the mobile ecosystem and tend to rely on visual cues, such as icon recognition and memorized gestures. Usability studies conducted in Low- and Middle-Income Countries (LMICs) have shown that users with low literacy typically comprehend only short and simple instructions, with audio, pictorial aids, and concrete examples significantly enhancing clarity~\cite{tuli2021actionable, barbareschi2018designing}. Additionally, hierarchical menus and technical jargon in app interfaces present considerable navigation challenges for these users~\cite{tuli2021actionable, barbareschi2018designing}. Furthermore, hyperlinked, lengthy privacy policies filled with technical jargon have been shown to increase cognitive burden and user frustration, causing users to experience privacy fatigue, issues that are further aggravated within populations with limited literacy~\cite{fatigue2021trust, felt2012android, barbareschi2018designing}. Gaps in financial and digital literacy compound these challenges, leaving users especially vulnerable to manipulation, misinformed consent, and exploitative data-use practices. These usability and comprehension barriers set the stage for exploitation by predatory apps that rely on opaque disclosures and manipulative design.

\subsection{Icon Comprehension and Visual Risk Cues} 
While mobile interfaces rely heavily on iconography to convey meaning, prior HCI research indicates that icons are often semantically opaque to users with low literacy. Barbareschi et al. note that abstract symbols often fail to translate across cultural and literacy barriers, with users preferring concrete, photorealistic representations over stylized glyphs~\cite{barbareschi2018designing}. For example, Tuli et al. found that common navigation icons require ``actionable'' contextualization to be understood by low-literate users in India, arguing that static symbols alone are insufficient for conveying complex states \cite{tuli2021actionable}.

In the specific context of security, ``visual risk communication'' becomes even more challenging. Standard warning icons (e.g., triangles or locks) are frequently misinterpreted. A lock icon, intended to symbolize ``security,'' is often perceived by low-literate users as ``access denied'' or a locked feature. To bridge this gap, recent work has moved toward visual metaphors and analogies. Zimmermann et al. showed that visual analogies can significantly enhance the comprehension of abstract privacy policies by mapping data flows to familiar physical concepts~\cite{Zimmermann2025_LetsGetVisual}. Similarly, Harbach et al. showed that risk perception increases when users are presented with tangible examples of their own data (e.g., their own photos) rather than generic warnings~\cite{harbach2014using}. Our work extends this by employing generative AI to create culturally grounded ``worst-case scenario'' visuals, moving beyond static icons to narrative-based risk communication.

\subsection{Mobile Permission Models \& AOFU Problem}
Mobile access permissions often operate on the Ask On First Use (AOFU) model, in which an app requests access to resources such as the microphone or camera the first time a related feature is used, and users are prompted to approve or deny these requests after reviewing them~\cite{wijesekera2017android}. However, Wijesekera et al.~\cite{wijesekera2017android} reported that 33\% of survey participants, despite having an average of five years of Android experience, were unaware that approving an AOFU prompt grants blanket permission for all future use cases. For users with low literacy, who often struggle to understand the implications of granting permissions in the first place, this number is likely to be substantially greater, and this model can result in applications obtaining virtually unrestricted and ongoing access to sensitive data. For predatory apps, the persistent access enabled by the AOFU model can be strategically exploited to harvest sensitive personal and behavioral data, which is then leveraged for coercive, deceptive, or financially exploitative practices. Section 2.5 examines three major categories of such applications: loan, gambling, and trading apps, and their documented patterns of abuse.

\vspace{-5pt}
\subsection{Predatory Financial Apps}
Chen et al.'s study on CULPRITWARE apps, which use deception, coercion, or other criminal techniques to generate profit, found that among the 843 apps analyzed, financial apps accounted for 42.24\% and gambling apps 30.96\%~\cite{chen2023lifting}. These apps often employ fake loan offers and cryptocurrency scams to deceive users, particularly those with low  digital literacy, into surrendering large sums of money. In addition to financial apps, gambling and trading apps often promise fast financial returns while concealing the volatility and risk involved. These apps have been found to use misleading interfaces and aggressive onboarding tactics (such as large starting sums of money), further endangering users with low literacy~\cite{acharya2024conning, bartoletti2021cryptocurrency, xia2020characterizing}. A common characteristic of all such apps was their significantly higher use of sensitive permissions compared to non-predatory counterparts~\cite{chen2023lifting}. Munyendo et al. documented that the most prevalent form of permission abuse by predatory financial apps was contacting users' contacts without consent when loans defaulted~\cite{munyendo2022desperate}. Multiple users reported that their relatives and friends were contacted and threatened and that they felt publicly shamed~\cite{munyendo2022desperate}. Despite the prevalence of such practices, recent work has rarely examined how these apps target users with low literacy specifically or whether interventions like simplified, LLM-generated summaries can help close the informed consent gap.

\vspace{-5pt}
\subsection{NLP and LLM-based Approaches to Informed Consent: Current Work and Gaps}
A number of automated approaches have sought to improve the accessibility and utility of privacy policies. Tools such as Polisis~\cite{harkous2018polisis} made use of neural-network-based classification techniques to segment and label privacy policy text, whereas PoliGraph~\cite{cui2022privacyglasses} restructured policy content into knowledge graphs. Both tools then enabled users to query specific practices such as data sharing with third parties. CLAUDETTE~\cite{lippi2019claudette} was used to detect unfair or non-compliant contractual clauses using supervised learning, while PolicyLint~\cite{xie2019policylint} used NLP techniques to identify outdated or contradictory statements in policies. More recent works have used transformer-based machine learning approaches to generate conversational, layperson-friendly summaries and have shown significant promise~\cite{pan2024clarify}. However, these systems have almost exclusively been evaluated with literate, digitally fluent populations in high-income countries, focusing largely on static policy documents while neglecting UI-level consent mechanisms such as permission prompts. Across these works, three critical gaps remain: (1) limited adaptation to the needs of users with low literacy, particularly in Low- and Middle-Income Countries (LMICs); (2) a lack of integration between policy text simplification and the interpretation of runtime permission requests; and (3) minimal empirical evaluation of whether LLM-generated summaries actually improve informed consent and contribute to prevention of financial exploitation in vulnerable populations.

Our work addresses these gaps by designing and evaluating an LLM-powered intervention that generates simplified summaries for both privacy policies and mobile app permissions, targeting Android users with low literacy in an LMIC context.

\section{Study Overview and Methodology}
To investigate the intersection of privacy literacy and predatory app behavior, we adopted a multi-phase mixed-methods approach. The study proceeded in three sequential phases:

PH1 (User Context): conducting semi-structured interviews with 34 factory workers to establish a baseline of permission awareness and privacy misconceptions \textbf{(RQ1)}; PH2 (App Analysis): performing static analysis of 50 popular financial apps to quantify the prevalence of over-permissioning and high-risk data requests \textbf{(RQ2)}; and PH3 (Intervention): Based on findings from PH1/PH2, we designed and evaluated LLM-generated interventions to measure their impact on user comprehension and risk sentiment \textbf{(RQ3)}.

\vspace{-5pt}
\subsection{Participants and Recruitment}
With IRB approval, we recruited 34 Android users (10 women, 24 men) from a manufacturing facility in Lahore, Punjab. The site was selected to access a demographic of blue-collar workers who frequently rely on smartphones but lack formal digital literacy training. The participants' ages ranged from 18 to 50 ($M=35.0, SD=9.3$) and their educational backgrounds varied significantly by gender, reflecting local disparities in schooling opportunities. Women reported fewer years of formal education ($M=2.0, SD=3.6, Median=0$) compared to men ($M=7.7, SD=3.2, Median=8$). Table~\ref{tab:age-edu} details these gender distributions. 

While prior work often categorizes such populations broadly as ``low-literate,'' we specifically screened for participants who self-identified as having difficulty reading or comprehending text-heavy mobile interfaces without assistance. As shown in Table~\ref{tab:age-edu}, while some participants possessed basic functional literacy in Urdu (their native language), the vast majority struggled with English, which is the default language of the Android OS and most privacy policies in this region. Therefore, we frame our population not as strictly ``illiterate,'' but as users with limited proficiency in English-dominant digital ecosystems. This distinction highlights that barriers to consent are often linguistic and structural rather than purely cognitive.

Participation was voluntary and conducted strictly during paid working hours to ensure no loss of wages for the workers. To avoid potential coercion associated with cash payments in a low-income workplace setting and to adhere to factory safety protocols regarding external financial transactions, we did not provide monetary compensation. Instead, we provided refreshments (juice and biscuits) as a culturally appropriate gesture of gratitude. This model was approved by the Institutional Review Board (IRB).

\begin{table}[h]
  \centering
  \small
  \caption{\small Participant distribution by age and education level ($N=34$).}
  \vspace{-10pt}
  \label{tab:age-edu}
  \begin{tabular}{lcc}
    \toprule
    \textbf{Category} & \textbf{Men - n (\%)} & \textbf{Women - n (\%)} \\
    \midrule
    \multicolumn{3}{l}{\textit{Age (years)}} \\
    \cmidrule(lr){1-3}
    18–25 &  6(17.6\%) & 1(2.9\%)  \\
    26–35 &  8(23.5\%) &  3(8.8\%)\\
    36–45 &  5(14.7\%) &  6 (17.6\%)\\
    46+   &  5(14.7\%) &  0(0\%)\\
    \addlinespace
    \multicolumn{3}{l}{\textit{Education completed (grades)}} \\
    \cmidrule(lr){1-3}
    0–5 (Primary)          & 7(20.5\%) & 8(23.5\%) \\
    6–8 (Middle)           & 6(17.6\%) & 1(2.9\%) \\
    9–14 (Secondary–Tertiary) & 11(32.4\%) & 1(2.9\%) \\
    \bottomrule
  \end{tabular}
  \raggedright\footnotesize
  \vspace{-10pt}
\end{table}

\subsection{PH1: Semi-Structured Interviews}
\begin{figure}[h]
  \centering
  \includegraphics[width=0.9\linewidth]{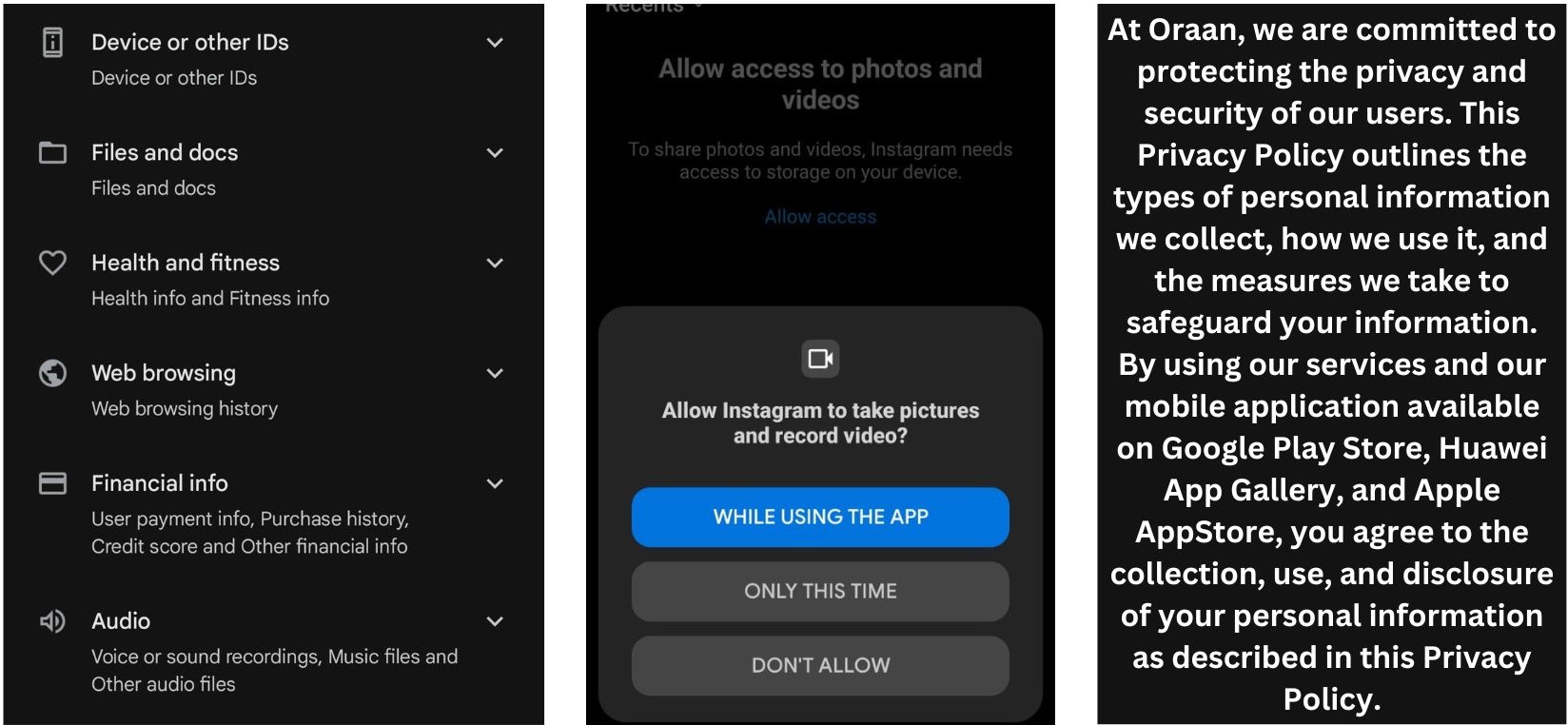}
  \vspace{-10pt}
  \caption{\small Play Store permissions (left), runtime permissions (center), and sample privacy policy (right).}
  \label{fig:permission_screenshots}
  \Description{The most commonly used apps in our participants' phones, according to their own perception}
\end{figure}

We conducted one-on-one, semi-structured interviews in Urdu (the local native language). To ensure cultural sensitivity and participant comfort, the interviews were gender-matched. The interviews lasted an average of 14 minutes and focused on four key themes: (1) perceived permission abuse, (2) awareness of privacy policies, (3) mental models of data responsibility, and (4) preferred disclosure channels. To aid recall, participants who could not remember permission prompts were shown a printed screenshot of a standard Android runtime permission dialog (Figure~\ref{fig:permission_screenshots}).

Following the interview, we conducted a structured debriefing to mitigate any anxiety caused by discussing privacy risks. Researchers explained the actual function of permissions (e.g., that ``Allow'' grants persistent access) and clarified misconceptions, such as the belief that deleting a photo from the gallery removes it from a server. Participants were also provided with actionable steps to revoke unnecessary permissions on their devices.

\vspace{-15pt}
\subsection{Qualitative Analysis and Positionality}
While data collection was qualitative, we adopted a quantitized qualitative approach for analysis. This involved combining descriptive statistics with reflexive thematic analysis ~\cite{braun2006thematic}. Interviews were transcribed into English in real time; however, culturally specific phrasing was transliterated to preserve meaning. and translated only during the coding phase. 
\begin{table*}[h]
  \centering
  \small
  \caption{\small Participants’ awareness and perceptions of app permissions and data responsibility}
  \vspace{-10pt}
  \label{tab: code_perm}
  \begin{tabular}{p{0.45\linewidth} p{0.5\linewidth}}
    \toprule
    \textbf{Question} & \textbf{Prominent Codes} \\
    \midrule
    Has an app ever asked you for permissions? Are you aware of any app that asks for permissions before it takes your data & 
    No (28), Never seen permissions being requested (7), Yes (5) \\
    \midrule
    Why do you think they take these permissions? & 
    I don't know (19), Functionality (2) \\
    \midrule
    Who do you think is responsible for the data, and if an app is able to access your personal data, who should be responsible for it? & 
    Personally Responsible (8), App Developer (9), Government (6) \\
    \midrule
    If so, how should application developers tell you about the data that they are accessing? & 
    Voice-based Communication (7), Short Summary in Urdu (6), Voice message (3), Transparency (3), Send Message (3) \\

    \bottomrule
  \end{tabular}
\end{table*}
\begin{table*}[h]
\centering
\small
\caption{\small Ability to read and understand a real finance app privacy policy (Oraan) by education tier.}
\vspace{-10pt}
\label{tab:policy_by_edu}
\begin{tabular}{p{0.44\linewidth} c c c c}
\toprule
\textbf{Reading/comprehension outcome} & \textbf{Grade (0-5)} & \textbf{Grade (6-8)} & \textbf{Grade (9-14)} & \textbf{Total (n, \%)}\\
\midrule
Cannot read at all                                 & 8 & 0 & 0 & 8 (23.5\%) \\
Can read letters only                              & 9 & 0 & 0 & 9 (26.5\%) \\
Can read small words                               & 0 & 2 & 0 & 2 (5.9\%) \\
Reads but cannot understand the policy             & 0 & 7 & 0 & 7 (20.6\%) \\
Reads and \emph{somewhat} understands the policy  & 0 & 6 & 0 & 6 (17.6\%) \\
Reads, understands, and can explain the policy     & 0 & 0 & 2 & 2 (5.9\%) \\
\midrule
\textbf{Column totals (n)}                         & 17 & 15 & 2 & \textbf{34 (100\%)} \\
\bottomrule
\end{tabular}
\end{table*}
One of the authors performed open coding to generate the initial codebook. To validate the reliability of these codes, a second independent researcher applied the codebook to a random subset of 20\% of the transcripts ($n=7$). We calculated Inter-Rater Reliability (IRR) using Cohen's Kappa, yielding a score of $\kappa = 0.82$, indicating strong agreement. Disagreements between coders were resolved through discussion to reach a consensus on the final themes (e.g., ``Privacy Fatigue'' or ``Proxy Reliance''). While this validation mitigates individual bias, we acknowledge that our interpretation remains influenced by our background. Therefore, we provide the following positionality statement: The analysis was conducted by researchers who are university-educated, privacy-literate, and fluent in both Urdu and English. We acknowledge that our position as ``digitally fluent'' may influence our interpretation. To mitigate this, we prioritized conceptual coherence over frequency, focusing on identifying structural barriers (e.g., language, UI design) rather than attributing findings to user deficits. In our reporting, participant counts ($n$) are included alongside key themes to contextualize the prevalence of mental models, though qualitative excerpts are used to deepen participants' reasoning.

\subsection {PH1 Findings: Awareness \& Mental Models}
\noindent \textbf{Perception of Permission Abuse in PFAs:}
While a majority of participants ($n=20$) reported no direct interaction with formal financial applications, their perceptions were heavily shaped by ``word-of-mouth'' narratives from their community ($n=5$). When asked about the risks of taking loans from apps, participants frequently cited the fear of excessive data access being used for coercion. For example, P14 noted the reputational threat caused by apps accessing contact lists: \enquote{...they [financial apps] call your cousins and annoy you.} Several participants ($n=5$) explicitly labeled data-collecting apps as ``fraud,`` with P11 stating: \enquote{I have heard they are fake... they take your data... we should not give them information.}

\noindent \textbf{Literacy and Policy Comprehension:}
As detailed in Table~\ref{tab:policy_by_edu}, comprehension of privacy policies was strongly correlated with education level. Participants with primary education ($0-5$ years) were largely unable to read the policy text at all ($n=8$) or could only recognize individual letters ($n=9$). Even among those with secondary education who could read, comprehension remained low; only two participants ($n=2$) were able to read, understand, and explain the policy. This confirms that English-only text policies serve as an effective barrier to consent for this demographic.

\noindent \textbf{Permission Awareness and Recall:}
Recall of permission prompts was notably low. Most participants ($n=28$) did not recall ever seeing an app ask for permission. Even after being shown a visual sample of a standard Android permission dialog (Figure~\ref{fig:permission_screenshots}, center), seven participants ($n=7$) maintained they had never seen such a screen, despite owning smartphones.

\noindent \textbf{Understanding of Data Access:}
When asked \textit{why} apps request permissions, the majority ($n=19$) responded, ``I don't know.'' Misconceptions were common; some participants believed permissions were requested ``to save users from fraud'' [P23], while others viewed them as a prelude to blackmail [P11]. Crucially, a subset of participants ($n=3$) fundamentally underestimated the technical capabilities of the device, with P4 asserting: \enquote{No, Photos can only end up in other hands if I send them to someone,} unaware that granting storage access allows background data exfiltration.

\noindent \textbf{Gatekeepers and Proxy Installers:}
Reliance on others was a recurring theme. Eight participants ($n=8$) reported relying on ``proxy installers'' (family members or shopkeepers who install apps on their behalf). Here, the permission dialogs typically appear during the initial setup performed by the proxy, meaning the primary user never encounters the consent decision. P33 noted: \enquote{The girl who downloaded things for me told me my data is protected,} illustrating how trust is transferred from the system to the human proxy.

\noindent \textbf{Data Access Responsibility:}
Participants expressed fragmented trust models regarding who is accountable for data protection. Responses were split between the app developers ($n=9$), the users themselves ($n=8$), and government regulators ($n=6$). A minority ($n=2$) invoked the Play Store's vetting process as a shield, with P9 stating: \enquote{Government, but everyone else; they said okay and then the app was released.} Two participants viewed data access as an inevitable cost of using the service, signaling privacy fatigue.

\noindent \textbf{Preferred Disclosure Channels:}
The vast majority ($n=30$) reported never being informed about how their data was protected or used. When asked how they \textit{should} be informed, preferences strongly favored voice-first support ($n=7$) and short Urdu summaries ($n=6$). Text-based disclosures were frequently cited as inaccessible; as P25 noted, \enquote{Call and ask since I can't message.} these informed the design of our audio-visual interventions.

\section{PH2: Static Analysis of Applications}

\subsection{Scope and Dataset Construction}
To answer \textbf{RQ2}, we curated a dataset of 50 Android applications by querying the Google Play Store in August 2024 (some apps may not be available on the Play Store anymore). We used targeted search terms in both English and Urdu, including: \textit{``loan,'' ``quick cash,'' ``betting,'' ``trading,'' ``invest,'' ``crypto,''} and their local language equivalents. These included apps from loan, gambling, trading, and other categories that have been previously identified in literature as posing increased risks of exploitative practices towards financially vulnerable populations \cite{chen2023lifting}.

Candidate applications were downloaded from the Finance and Casino categories. Regional settings were configured to Pakistan to ensure relevance to the study population. We included apps if they (1) offered services related to digital loans, gambling/betting, or financial trading; (2) had $\ge$ 10,000 installs to ensure a baseline of user reach; and (3) were free to download (to align with common user acquisition patterns in contexts with limited literacy).

We deliberately curated this dataset to be problem-driven rather than volume-driven. Our goal was not to provide a market-wide census but to analyze the specific class of applications that intersect financial risk with data privacy. The final dataset included financial applications spanning loans, BNPL (buy-now-pay-later), earned wage access, banking, remittances, crypto, and government payments, in addition to non-financial comparators such as lifestyle, gaming, and file-sharing apps. This ensured that we captured both (1) the core set of apps where privacy and financial harms are concentrated and (2) secondary contexts that shed light on broader permission-request practices.

Unlike large-scale scraping studies that focus on statistical prevalence, our analysis required deep, contextual inspection of user comprehension and consent flows. Such fine-grained qualitative work is methodologically incompatible with very large samples. Therefore, a smaller but strategically chosen dataset was the most valid approach for answering our research questions. We acknowledge that our dataset is not exhaustive; apps were excluded if they were niche services with low adoption or duplicate offerings. However, given that our findings generalize across the major functional categories and risk profiles, expanding the dataset would yield diminishing returns without substantively altering the conclusions.

Loan applications dominated the dataset because they were the most frequently returned results for finance-related queries in the Pakistan-specific Play Store and are widely advertised through social media targeting low-income users. Prior work has shown that such apps often request permissions far beyond their functionality needs and may engage in coercive debt recovery tactics~\cite{munyendo2022desperate}. This dataset captures the app categories most likely to intersect financial risk with invasive data practices in contexts where users with limited literacy have limited capacity to evaluate privacy risks.

\subsection{Analysis Methodology}
We statically analyzed the 50 selected applications using the Mobile Security Framework (MobSF)~\cite{abraham2023mobsf} to extract declared permissions. We mapped these against the Android Developer Documentation to categorize them into two groups: (1) \textbf{Normal permissions} (e.g., internet access) that are automatically granted and (2) \textbf{dangerous permissions} (e.g., contacts, location, microphone) that require explicit runtime consent due to their privacy impact \cite{android-permissions-overview}.

While MobSF classifies permissions based on potential misuse, we adhered strictly to Android's official classification of ``dangerous permissions'' to highlight the potential abuse of sensitive user data. We defined ``over-permissioning'' by mapping an app's claimed core functionality to its minimum required permissions. An app was marked as over-permissioned if it requested more than one dangerous permission unrelated to its functionality. This threshold was selected to avoid flagging edge cases where a single unrelated permission might be justified by secondary but legitimate features.

\subsection{Key Findings}
\begin{figure}[hbt]
  \centering
  \includegraphics[width=\linewidth]{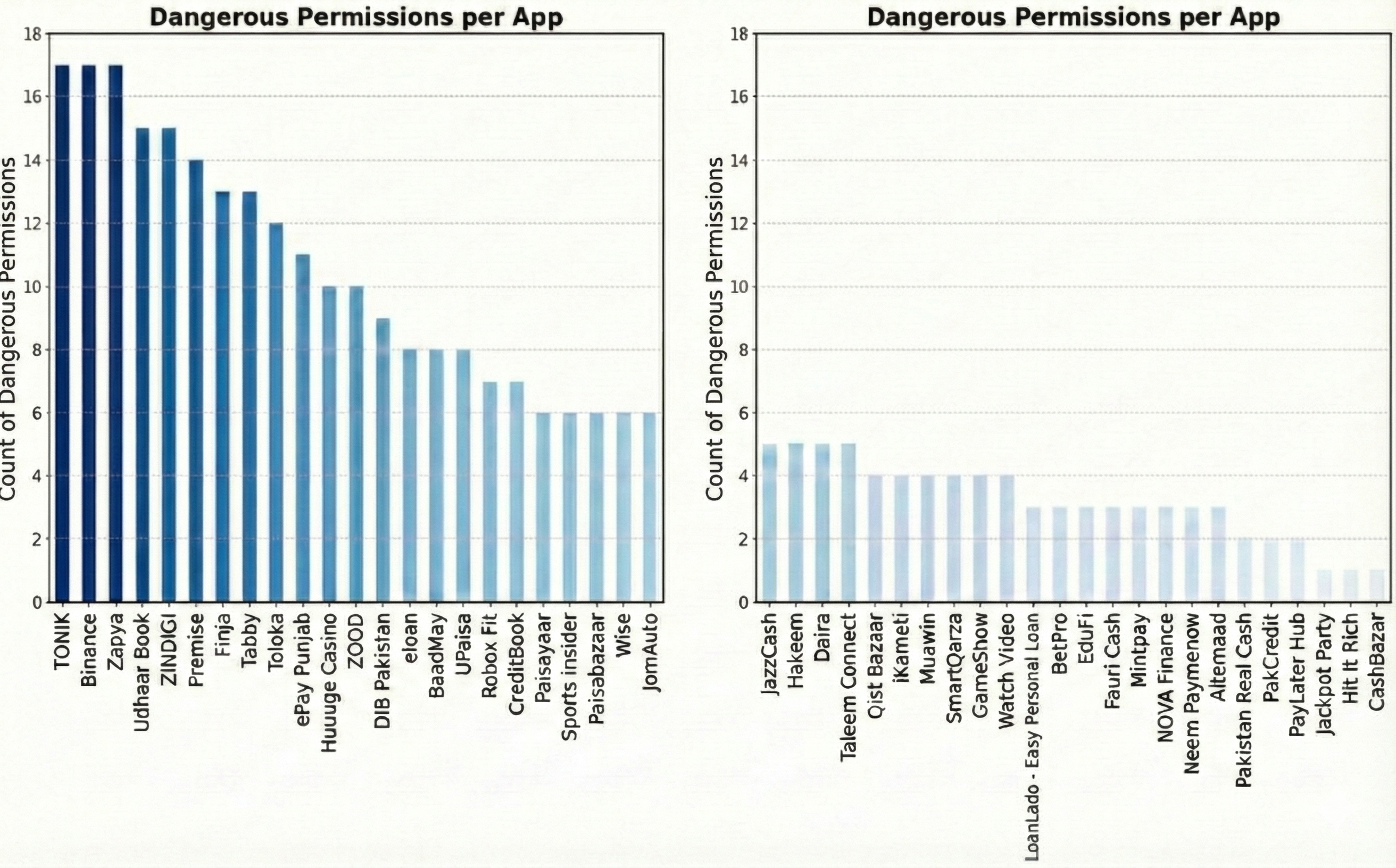}
  \vspace{-20pt}
  \caption{\small Number of MobSF-classified dangerous permissions across 50 shortlisted applications.}
  \label{fig:mobsf_results}
  \Description{Number of MobSF classified dangerous permissions across 50 shortlisted applications.}
\end{figure}

\begin{figure}[hbt]
  \centering
  \includegraphics[width=\linewidth]{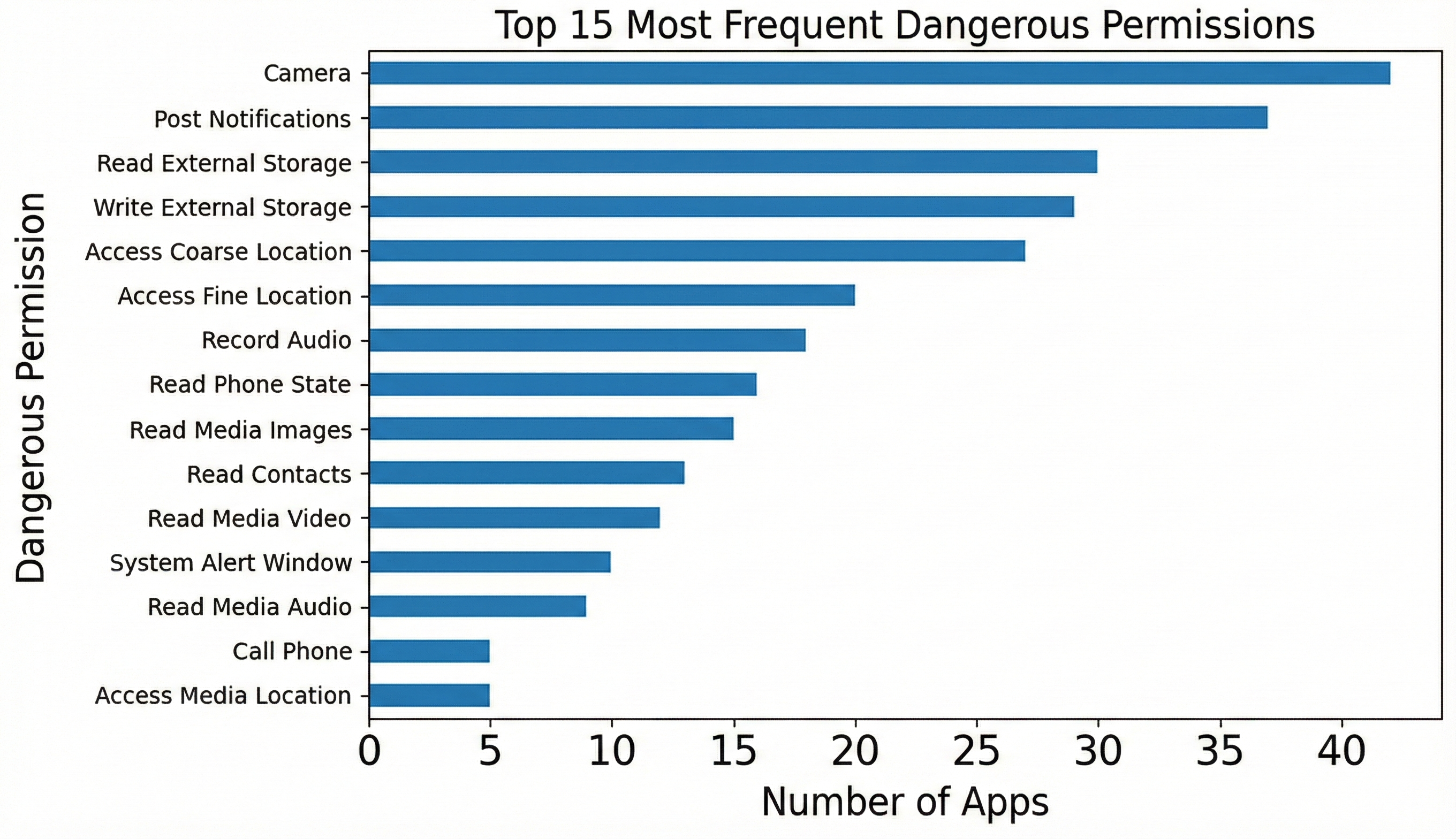}
  \vspace{-20pt}
  \caption{\small Top 15 most frequent permissions across our app suite.}
  \label{fig:perm_freq}
  \Description{Bar chart showing the top 15 most frequent dangerous permissions. Camera is the highest, followed by Post Notifications and Read External Storage.}
  \vspace{-10pt}
\end{figure}

Figure~\ref{fig:mobsf_results} illustrates the distribution of dangerous permissions across the dataset. Our analysis yielded two primary findings:
(1) \textbf{Prevalence of Over-Permissioning:} 80\% (40/50) of the apps requested permissions beyond those required for their primary functionality. We observed that \texttt{ACCESS\_COARSE\_LOCATION} was the most prevalent unnecessary request, though several apps also requested \texttt{ACCESS\_FINE\_LOCATION}. Notably, microloan apps frequently requested access to call logs and SMS history, data that is not strictly essential for loan processing but is often leveraged for user profiling or social pressure. Over-permissioning was most common among gambling/loan apps; (2) \textbf{Extensive Access to Sensitive Data:} As shown in Figure~\ref{fig:perm_freq}, a significant subset of applications requested high-risk permissions granting access to personal identifiers and media. For instance, gambling applications frequently requested GPS access under vague justifications like ``improving user experience.'' In the absence of accessible privacy policies, these requests effectively expose users' geolocation and social graphs to third parties without informed consent.
These patterns are particularly concerning for the financially vulnerable population in our study, who (as shown in PH1) often lack the ability to interpret the implications of these technical requests.

\vspace{-5pt}

\section{PH3: Intervention Design and Evaluation}

\subsection{Rationale and Pipeline}
\begin{figure}[t]
  \centering
  \includegraphics[width=0.9\linewidth]{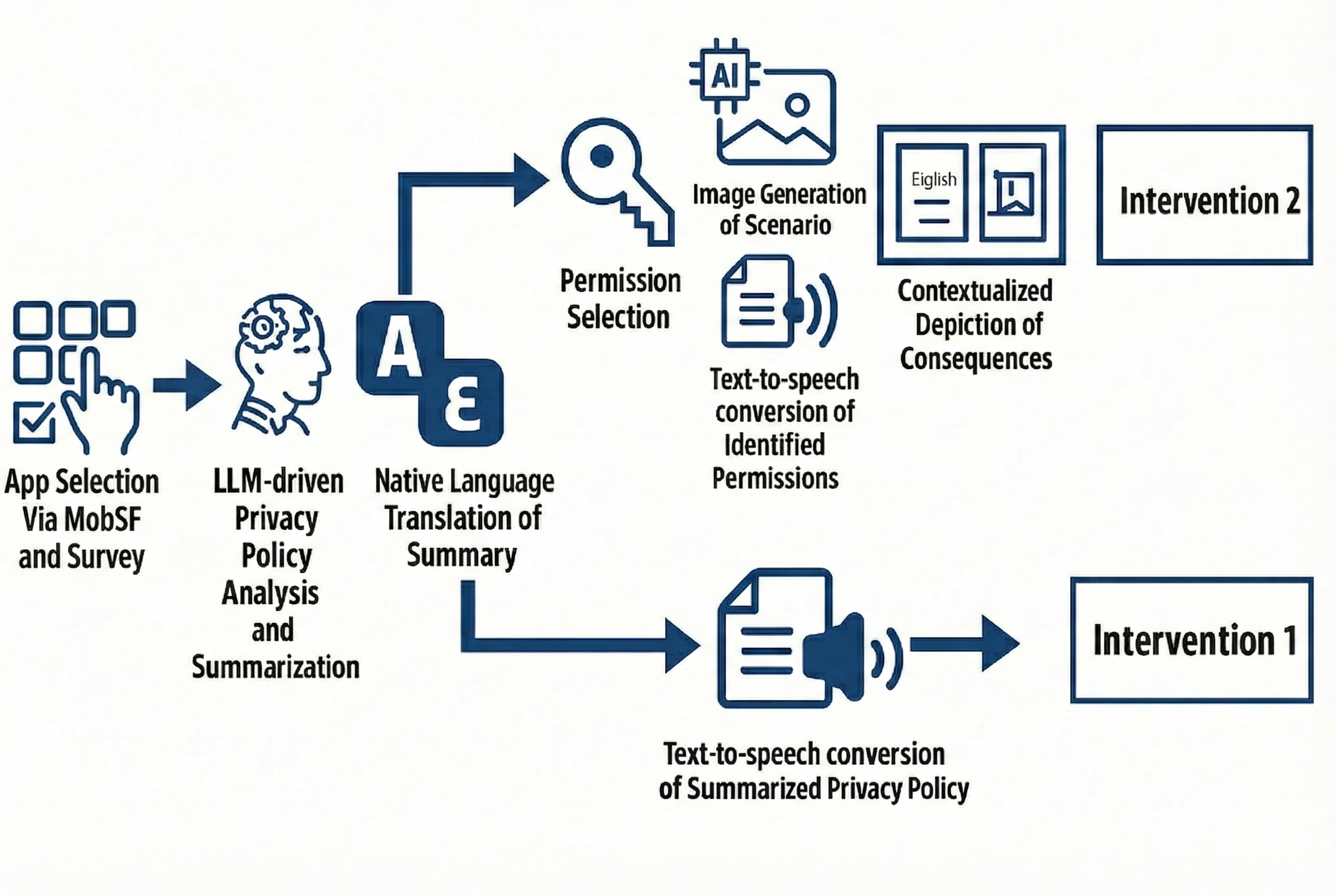}
  \vspace{-20pt}
  \caption{\small Stepwise pipeline for our privacy policy interventions}
  \label{fig:interventions_pipeline}
  \Description{We summarised the privacy policies and applications used by the participants in our intervention, and narrated the summary, both using LLM's}
  \vspace{-10pt}
\end{figure}

For users with limited text proficiency, the findings from PH2 reveal a compounded risk: permissions are technically opaque, and privacy policies are linguistically inaccessible. To address \textbf{RQ3}, we designed an automated pipeline to transform these complex disclosures into plain-language audio summaries and culturally grounded visual cues. We implemented a four-step generation pipeline (Figure~\ref{fig:interventions_pipeline}):
(1) \textbf{Policy Extraction:} We scraped privacy policies and declared permissions for the 50 applications in our dataset; (2) \textbf{Summarization (LLM):} Using Google's \texttt{Gemini 1.5 Flash}, we generated domain-specific summaries focusing on three key elements: (a) data collected, (b) usage purpose, and (c) potential risks. We utilized \texttt{gemma-3-1b-it} initially but encountered context-window limitations with lengthy legal texts, leading us to select Gemini 1.5 Flash for its scalability; (3) \textbf{Translation:} Summaries were translated into Urdu using a persona-based prompt to ensure colloquial relevance rather than formal legal translation; and (4) \textbf{Audio Synthesis:} Translated text was converted to speech using the Google Text-to-Speech (\texttt{gTTS}) library to bypass reading barriers entirely. To ensure reproducibility, the used prompts were:

\begin{itemize}[noitemsep,leftmargin=.1in]
    \item \textbf{Summarization Prompt:} ``Consider yourself an expert summarizer. Summarize the policy text into a maximum of 250 words. Keep the wording such that it feels spoken by a representative of the app. Keep the text simple and easy to understand, as the intended audience is lowly literate and might face difficulty in general reading and understanding.''
    \item \textbf{Translation Prompt:} ``I want to share this info with Pakistanis with low literacy who only know Urdu. You are an expert English-to-Urdu translator. Generate a translation of the summary that is simple, easy to understand, and accurate to the meaning.'' 
\end{itemize}
\vspace{5pt}
 LLMs were selected over human experts due to their scalability and ability to generate summaries rapidly across a large number of applications, and in this context, the primary objective was to enhance comprehension for users with low literacy rather than to achieve exhaustive legal precision. Moreover, LLMs were preferred over Small Language Models (SLMs) for summary generation, as SLMs frequently encounter context limitations when processing lengthy privacy policies, and this constraint reduces their practicality for such tasks. In our own testing to summarize a given privacy policy using \texttt{gemma-3-1b-it}, the model immediately reached its rate limit, and for these reasons, LLM-generated outputs offered a pragmatic and efficient choice. 

\subsection{Validation and Legal Review}
To mitigate the risk of LLM hallucinations or critical omissions, we conducted a validity check. Two independent legal professionals specializing in digital privacy law reviewed a random sample of 20\% of the generated summaries ($n=10$). Reviewers coded for (1) \textbf{Critical Hallucinations} (inventing data practices) and (2) \textbf{Omissions} (failing to mention a high-risk permission). The review found zero instances of critical hallucinations, though 2 summaries required minor edits to clarify data-sharing clauses. This ensures the simplified language maintains legal accuracy.

\subsection{Visual Risk Communication Design}
Our second intervention augmented the audio summaries with contextual visual cues. Leveraging work on low-literacy design~\cite{dowse2010developing}, we map abstract permissions to concrete, ``worst-case'' scenarios.

\noindent \textbf{Design Tensions and Paranoia:}
We intentionally designed these visuals to be ``alarmist'' to counteract the habituation and ``banner blindness'' observed in P1. For example, \texttt{ACCESS\_FINE\_LOCATION} was not depicted as a map pin but as a stranger tracking a home (Figure~\ref{fig:loc}). Similarly, \texttt{CAMERA} access was depicted as a recording of private family moments (Figure~\ref{fig:contact}).

Given the predatory nature of the specific apps analyzed (e.g., loan sharks blackmailing families), we prioritized \textit{alertness} over \textit{neutrality}. The images were generated using the \texttt{gpt-4-0613} model with the prompt:
``Consider yourself an expert in designing simple, clear, and culturally appropriate visual aids for Pakistani users with low literacy... The visuals should: (1) Use familiar everyday objects... (2) Be simple and uncluttered... (3) Use clear actions... (4) Highlight warnings... (5) Be inclusive and culturally sensitive.''

\subsection{Evaluation Procedure}
We evaluated the interventions using a controlled within-subjects experiment with the 34 participants recruited in PH1. The procedure followed three steps: (1) \textbf{Baseline Assessment:} Participants were shown the standard Play Store listing of a financial app and asked to rate their confidence in installing it (Scale: Confident, Neutral, Cautious, Distrustful), (2) \textbf{Intervention 1 (Audio/Text):} Participants listened to the LLM-generated Urdu summary. They then re-evaluated their confidence, and (3) \textbf{Intervention 2 (Visuals):} Participants were shown the generated ``risk visuals'' alongside the audio and provided a final confidence rating.

\subsection{Results}
\subsubsection{Intervention 1: LLM-Generated Summaries (Text + Audio)}
When presented with simplified, audio-supported summaries, participants' trust in apps decreased substantially. As shown in Table \ref{tab:intervention1}, women shifted from 73\% ``Confident'' to 42\% ``Cautious,'' while men shifted from 88\% ``Confident'' to 86\% ``Cautious.'' This suggests that providing policy content in accessible language alone can meaningfully alter user perceptions.

\vspace{-10pt}
\begin{table}[!ht]
  \centering
  \caption{\small Intervention 1 results as separated by gender.}
  \vspace{-10pt}
  \footnotesize
  \label{tab:intervention1}
  \begin{tabularx}{\columnwidth}{l X X}
    \toprule
    Gender & Previous Impression & Impression after Intervention \\
    \midrule
    Men & 88\% Confident, 21\% Distrustful & 86\% Cautious, 7\% Confident, 7\% Neutral \\
    \addlinespace 
    Women & 73\% Confident, 17\% Neutral & 42\% Cautious, 45\% Neutral, 13\% Confident \\
    \bottomrule
  \end{tabularx}
  \vspace{-15pt}
\end{table}

\subsubsection{Intervention 2: Visual Cues}
The addition of visual cues produced even stronger effects (Table \ref{tab:intervention}). Men shifted to 94\% ``Cautious,'' while women shifted to 92\% ``Cautious.'' Qualitative feedback indicated that participants found the visuals ``memorable'' and ``more believable'' than text alone. Participants reported that images helped them connect permissions (e.g., GPS) to concrete privacy risks.

\begin{table}[!ht]
\vspace{-10pt}
  \centering
  \caption{\small Intervention 2 results separated by gender.}
  \vspace{-13pt}
  \footnotesize
  \label{tab:intervention}
  \begin{tabularx}{\columnwidth}{l X X}
    \toprule
    Gender & Previous Impression & Impression after Intervention \\
    \midrule
    Men & 88\% Confident, 21\% Distrustful & 94\% Cautious, 6\% Neutral \\
    \addlinespace 
    Women & 73\% Confident, 17\% Neutral & 92\% Cautious, 5\% Confident \\
    \bottomrule
  \end{tabularx}
\end{table}

\vspace{-15pt}
\subsubsection{Comparative Impact}
Overall, Intervention 2 led to higher increases in caution than Intervention 1 across both genders. These findings indicate that visual augmentation, when combined with simplified language and native audio, can significantly improve risk awareness in users with low literacy.

\section{Discussion}

\subsection{Beyond the Deficit Model}
Our findings demonstrate that the factory workers interviewed in this study faced significant barriers to making informed decisions, reinforcing concerns raised in prior work on mobile consent mechanisms~\cite{wijesekera2017android, munyendo2022desperate, chen2023lifting}. This challenges the prevailing ``deficit model'' of informed consent, which assumes users simply need ``simpler text.'' Instead, our study reveals that consent in this demographic is structurally undermined by a power imbalance. The reliance on ``proxy installers'' (gatekeepers) and the inability to decode English-dominant interfaces strips users of agency long before they encounter a permission prompt.

By deploying LLM-generated audio and ``worst-case'' visuals, our intervention does not merely improve comprehension. It attempts to disrupt this power asymmetry by equipping users with the same risk intelligence possessed by the developers. This supports earlier research showing that multimodal consent cues improve comprehension for users with low literacy~\cite{dowse2010developing} and extends it by demonstrating the feasibility of using LLMs (specifically \texttt{gpt-4-0613} and \texttt{gemini-1.5-flash}) to produce these cues at scale.

\vspace{-5pt}
\subsection{Designing for Redundancy \& Consequence}
The success of our intervention lies in \textit{redundancy}. Prior research notes that standard icons are often semantically opaque to users with low literacy~\cite{tuli2021actionable}. Our design bypasses this dependency: by combining simplified text, contextual imagery, and voice narration, we ensure the message survives even if one modality fails.

Furthermore, we propose a shift toward \textbf{Consequence-Based Design}. Mobile operating systems should move beyond abstract icons (e.g., a generic ``Location'' pin), which fail to convey harm. Our results suggest that users react to \textbf{narrative consequences}. We recommend standardized, OS-level visualizations that depict the data \textit{leaving} the device (e.g., a visual of a map being sent to a server) rather than just the access itself.

\vspace{-5pt}
\subsection{Regulatory Blind Spots and Policy Gaps}
Our static analysis reveals that exploitative practices are not limited to high-visibility platforms but are widespread in the ``long tail'' of financial applications. While Google revised its Play Store Personal Loans policy in 2023 to restrict sensitive data access~\cite{techcrunch2023googleloanapps,googleplay2023loanpolicy} and mandated the photo picker for post-2024 apps~\cite{googleplay2024loanpolicyupdate}, these mechanisms focus on apps with lending as a \textit{primary} function.

This creates a regulatory blind spot for \textbf{hybrid applications} where lending is a secondary feature (e.g., e-commerce or lifestyle apps), as well as the \texttt{Culpritware}~\cite{chen2023lifting} family of apps distributed via side-loading. Our proposed LLM-based intervention addresses this by being inherently scalable to \textit{any} APK, regardless of its category or channel. We recommend that regulatory bodies in South Asia (e.g., SECP in Pakistan, RBI in India) mandate a \textbf{``Voice-First Disclosure''} for all high-risk apps, requiring a compulsory audio summary of critical data rights prior to financial disbursement.

\vspace{-5pt}
\subsection{Limitations}
Several limitations should be noted when interpreting our findings:
(1) \textbf{Participant sample and demographics:} Our user study involved a limited number of participants drawn from a specific urban-peri-urban region in Pakistan. While we controlled for literacy level, we did not systematically measure language proficiency, digital experience, or socio-economic variation, all of which may influence comprehension and decision-making. Owing to the skew amongst participants in terms of gender, our findings may not generalize to all populations with limited literacy, nor to other regions within the `developing world' or `LMICs' that may have different cultural or linguistic dynamics. (2) \textbf{Cultural and contextual specificity:} Both the policy summaries and the visual cues were designed for a Pakistani LMIC context, leveraging Urdu/Punjabi translations and locally resonant imagery. While this supports ecological validity for our target population, it may limit applicability to other cultural and linguistic settings without adaptation. (3) \textbf{Recall Bias:} Several participants reported ``never seeing'' permission prompts. Given the prevalence of proxy installers, this likely reflects a lack of recall rather than the absence of prompts. (4) \textbf{Visual Piloting and Paranoia:} We acknowledge an ethical tension in our design: while effective, our ``worst-case'' visuals (e.g., recording family members) were not separately piloted for psychological impact. Such alarmist imagery risks inducing excessive paranoia, potentially deterring users from legitimate financial inclusion tools. Future work must calibrate this ``fear appeal'' to balance caution with trust.

These limitations notwithstanding, the study offers initial empirical evidence that automated, multi-modal consent interventions can markedly improve comprehension for users with low literacy.

\bibliographystyle{ACM-Reference-Format}
\balance
\bibliography{sample-base}

\appendix

\onecolumn

\section{Appendices}
\label{appndx}
\begin{figure}[hbt] 
  \centering
  \begin{subfigure}[b]{0.45\textwidth}
    \centering
    \includegraphics[width=\linewidth]{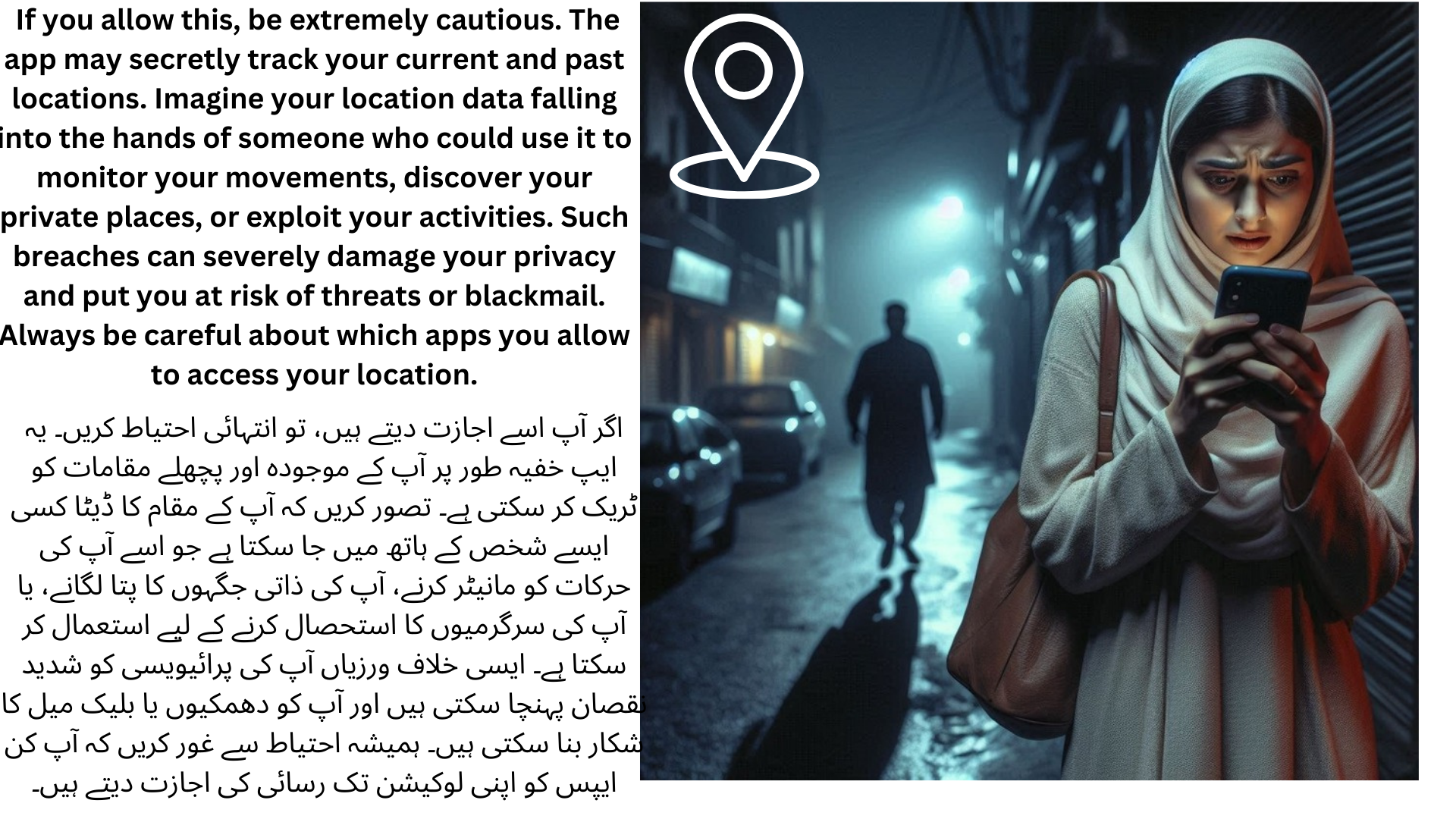}
    \caption{Location permission intervention.} 
    \label{fig:loc}
  \end{subfigure}
  \hfill
  \begin{subfigure}[b]{0.45\textwidth}
    \centering
    \includegraphics[width=\linewidth]{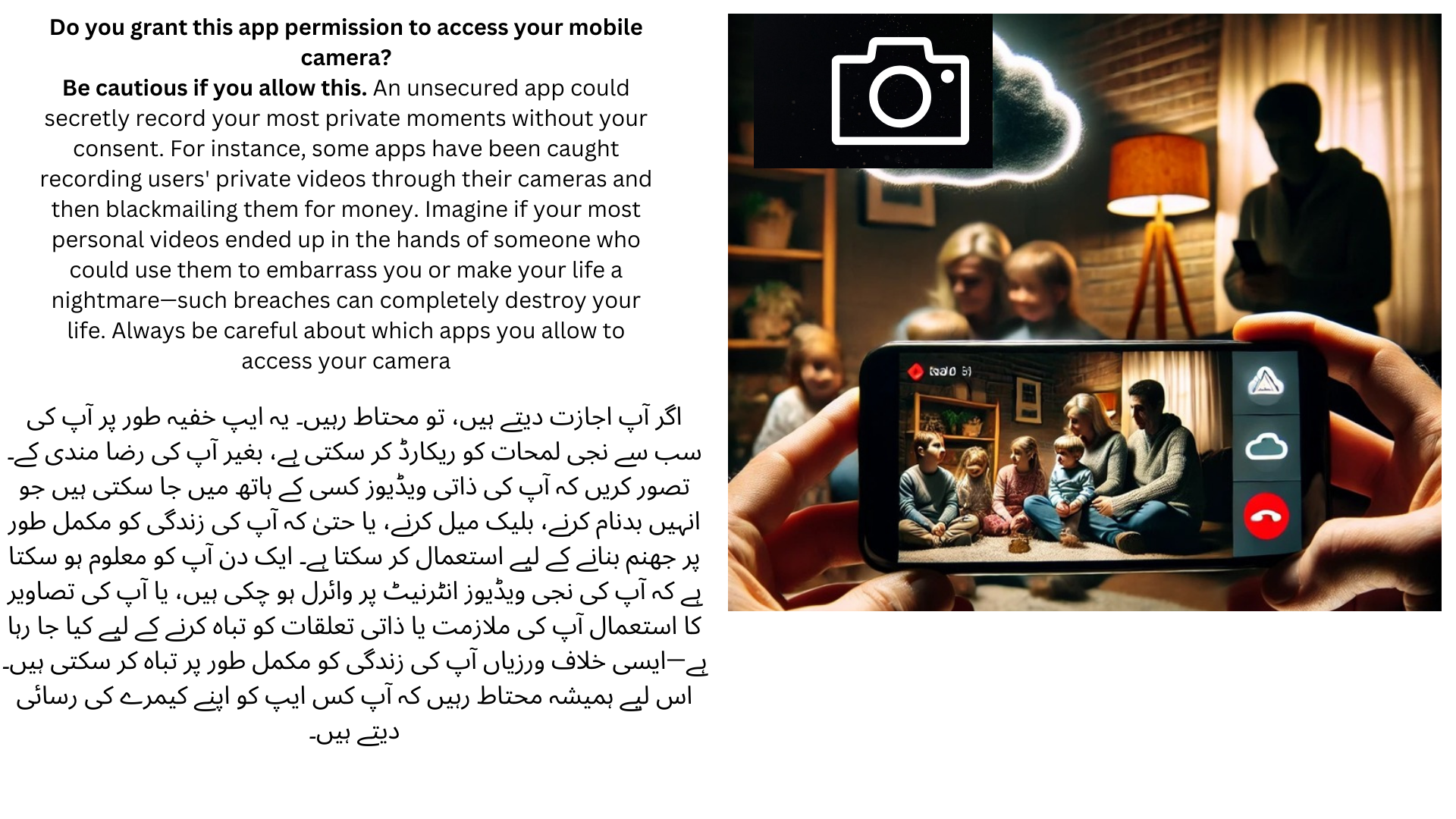}
    \caption{Camera permission intervention.} 
    \label{fig:contact}
  \end{subfigure}
  \caption{The intervention designs: Location (Left) and Camera (Right).}
  \label{fig:both_interventions}
\end{figure}

\begin{table}[!hbt]
\centering
\caption{Applications in the test suite are grouped by broad category and subcategory (n=50).}
\label{tab:apps_by_category}
\footnotesize
\begin{tabular}{@{}p{0.16\textwidth} p{0.22\textwidth} p{0.54\textwidth} r@{}}
\toprule
Broad category & Subcategory & Applications \\
\midrule
Finance & Loan (NBFC) - 17 apps & Hakeem, Paisayaar-Easy Safe Cash Loan, Tabby, JomAuto, Mintpay, JazzCash, Daira, Eloan, Paisabazaar, SmartQarza, PayLaterHub, Money Tap, Aitemad, Fauri Cash, Loan Lado, Pak Credit, CashBazar\\
& BNPL \& EWA - 10 apps & Edufi, BaadMay, QistBazaar, Abhi - Your Salary Now!, Zood, CreditBook, Muawin, Taleem Connect, Neem Paymenow, Nova Finance\\
& Banking - 3 apps  & Zindigi, Tonik - Fast Loans \& Deposits, DIB \\
 & Funds Transfer - 3 apps & Upaisa, Wise, Finja \\
 & Crypto - 1 app & Binance\\
 & Government Payment - 1 app & Epay \\
\midrule
Business & Management App - 2 apps & iKameti, Udhaar Book: Digi Khata \& Earn\\
\midrule
Lifestyle & Earn Cash - 6 apps & Toloka: Earn online, Robox Fit - Walk \& Earn, Premise - Earn Money for Tasks, Repocket - Make Money Daily, Pakistan Real Cash, WatchVideo \\
 & Game - 4 apps & Huuuge Casino 777 Slots Games, Hit it Rich! Casino Slots Game, Jackpot Party Casino Slots, Gameshow\\
 & Sports - 2 apps & SI-Betting tips, BetPro App\\
\midrule
Tools & File Sharing - 1 app & Zapya\\
\bottomrule
\end{tabular}
\end{table}

{\footnotesize
\begin{longtable}{P{0.34\linewidth} P{0.46\linewidth} P{0.16\linewidth}}
\caption{Interview questions and resulting codes/themes (non-mutually exclusive). Counts are participants (n/34).}
\label{tab:codebook-all}\\
\toprule
\textbf{Question} & \textbf{Codes / themes (n)} & \textbf{Quote}\\
\midrule
\endfirsthead
\toprule
\textbf{Question} & \textbf{Codes / themes (n)} & \textbf{Quote}\\
\midrule
\endhead
\midrule
\multicolumn{3}{r}{\emph{Continued on next page}}\\
\midrule
\endfoot
\bottomrule
\endlastfoot

\multicolumn{3}{l}{\textbf{Section 1: Demographic and Background Information}}\\
Do you have a smartphone? & Yes (33); No (1) & -\\
Do you live in a \emph{rural} or \emph{urban} area? & Urban (19); Rural (15) & -\\
\hline
What is your age? & Attribute (see demographics table) & -\\
\hline
What is your highest degree of education? & Attribute (see demographics table) & -\\

How comfortable are you in understanding and writing English? \newline (Follow-up: Participant reads Oraan’s Privacy Policy for comprehension check.) &
Cannot Read at all (8); Can read letters only (9); Can read small words (2); Read, Cant Understand (7); Read and Somewhat Understand PP (6); Read and Understand and Explain PP (2) & -\\
\hline

\addlinespace[0.6em]
\multicolumn{3}{l}{\textbf{Section 2: Smartphone Usage and App Familiarity}}\\
\hline
What is the name and model of your smartphone? (If unsure, ask to see the device.) & Descriptive (device brand/model) & -\\
\hline
What apps do you use regularly on your phone? &
Whatsapp (25); Facebook, Instagram (21); Tiktok (18); Games (12); Finance Apps (9); YouTube (10); Snapchat (4); Tamasha (2) & -\\
\hline
How do you download apps? (play store or otherwise) &
Playstore (13); Ambiguous (1); Voice Command (7); Trusted person downloads (8); Search Bar (3); word of mouth (3); Grey Tick/ (*) sign -> Okay to download (2) & ``My husband downloads for me.'' [P32]\\
\hline
Are there any gaming applications? &
Yes (18); No (12); Ludo (11); Only for kids (1); No use due to low literacy (1) & -\\
\hline
Are any of these card games and prediction games? &
No (27); Yes (3) & -\\
\hline

\addlinespace[0.6em]
\multicolumn{3}{l}{\textbf{Section 3: Financial and Loan Application Experience}}\\
\hline
Do you know of any financial applications? Are you currently using any or have used any in the past? &
JazzCash (18); EasyPaisa (13); Don't know (9); Uses (3); Mobile Wallets (2); Banking App (1); BInance, KUCoin, TradingView, MarketCap (1); Paypal (1); Believes in Trading Play \& Win (1); Not literate to play trading games (1); Not literate to use finance apps (4); Knows but doesn't use (2); Trusted user made Fin Acc (1) & ''Jazzcash, don't know about Easypaisa, don't know about loan apps.`` [P12]\\
\hline
Do you know of any loan apps? &
Yes (8); No (17); Took Loan (5); Emergency (1); Interest Increases after limit (3); Weekly Charges added on (3); Don't Annoy (1); Haven't Tried (3); Message (1 Call) received on taking Loans + you won (4); Don't Want to use due to High Charges/Interest (2); Not capable of taking loan (3); Fake perception (1) & ``I've heard of high penalties for taking loans, so I never took them.'' [P3]\\
\hline
How has your experience been while using loans? &
No problem / Good / Fine Exp (4); Receive Message on Loan (2); Play and Earn (1); No Exp (18); Loan from Trusted Member (Family/Friend) (2) & ``Had a good experience, no issue.'' [P10]\\

If someone is in financial need, how do you feel about them taking loans. &
Learn (Know Interest/Terms and Conditions) (2); Take loan in dire condition (6); Take from someone you trust (7); Fake/Fraud Perception (6); Extra Charges/Interest (4); More Accessible in App (1); Trust in Jazzcash \& EasyPaisa (3); Yes (Application) (18); No (Application) (13); No idea (1); Conflict (2) & ``No they should not, but they would get in more difficulty; it is better to take from someone you know.'' [P8]\\
\hline
What prerequisite information does the application need to give you the loan? &
CNIC - etc (18); no extra info apart from Jazzcash/EasyPaisa account (3); extra unnecessary information (7); interest/sood (1); No idea (6); External Help (1) & ``only Easypaisa profile.''\\
\hline
What do you think happens when you don’t pay the loan? &
FIR/Police (9); Fraud (3); Increased Charges to be paid (6); Block Account (2); Close Account (1); Sim Blocked (2); Come to house (4); Bother + Fight + Bully (10) & ``Can block the number and can hack data and use it against me.'' [P14]\\
\hline
Why do you use this mobile application? &
Trust due to Others Using it (2); Emergency (2); Financial needs (Loan) (1); don't get loan (1); Only Transactions (5); To buy (costly things) (2); Min Phone Usage (1) & ``Send and  receive money.'' [P19]\\
\hline
Have you stopped using any mobile loan application? (Give additional details?) &
Use it for Transactions (1); Could not figure out how to use it (3); Transactions (1) & ``Not able to.'' [P27]\\
\hline
Have you ever used a loan or gambling app? &
Yes (2); No (22) & -\\
\hline

\addlinespace[0.6em]
\multicolumn{3}{l}{\textbf{Section 4: Data Privacy and Permissions Awareness}}\\
\hline
Has an app ever asked you for permissions? Are you aware of any app that asks for permissions before it takes your data &
Yes (5); No (28); Terms \& Conditions-> Long - Don't read (1); Don't know what they do with permissions (1); Don't care (because of other important things \& they don't think they have sensitive information in phone) (2); Word of Mouth - Trust Application (1); Never seen permissions being requested (7); Somewhat understanding/recall of Permissions (3); Inability to understand permission (1); Trusted Person (1) & ``No, but my nephew installed it, so I don't use it.'' [P28]\\

Why do you think they take these permissions? &
I don't know (19); I don't care (1); Precaution against fraud users (1); Never thought, why would an app take permission (1); Identification (1); Blackmail Requirement (1); Not literate enough to know why permissions are taken (1); Pushes the idea of an application asking for permissions (1); Functionality (3); Ludo is taking Phone Number (1) & ``Whatsapp for contact application usage.'' [P19]\\
\hline
Who do you think is responsible for the data, and if an app is able to access your personal data, who should be responsible for it? &
Personally responsible for their data (8); the app developer is responsible for their data (9); worry about data (2); government (6); Play Store Authority (2); the one who takes info and uses info (3); app owner (1); the person who manages the phone (2); agencies have access to data in the phone. (1); App accessing phone and gallery data is worrisome (1) & ``My brother; app is responsible.'' [P4]\\
\hline
Are you aware of any rules and regulations that protect your personal information? &
Yes (2); No (28); Believes the only entity for sharing data is him alone (2); Believes there should be rules and regulations (2); PTA rules (1); the government should not allow (1); Fire Safety (Rescue 1122) (1); Thinks only God can protect (1); Messages sent by Govt and Bank (1) & ``There should be one,  but I haven't heard one.'' [P31]\\
\hline
What is the most important data that you have on your phone? &
Gallery (21); Contacts (6); Messages (Whatsapp) (6); Family (7); Personal Phone Number (4); Voice Notes (2); Google ID (1); Bank info (1); Facebook ID (1); Nothing (1); Jazzcash (1) & ``Childrens Photos.'' [P32]\\
\hline
Have you been informed by an app or device on how they protect your data? &
Yes (3); No (30); Wont give permission if requires access to phone data (1); The person who they trust said data is protected (1) & ``No they never have.'' [P13]\\
\hline
Do you think the applications developer is responsible for your data protection? &
Yes (27); No (4); Trust in Developer (1); No trust in Developer (1) & -\\

If so, how should they tell you about the data that they are accessing? &
Short, concise summary in Urdu (6); Be clear if they are doing something fishy - say ``We are fishy'' (1); Transparency needed (3); Blackmail (2); Accessing data is problematic (1); Send Message (3); Knows about cybercrime (1); They access data for verification purposes (1); Helpline/Call/Customer Support (7); Don't access data - its private (2); don't access - until I do something wrong/inappropriate (1); don't ask repeatedly for data - fatigue (1); the government should be more active (1); voice message (3) & ``They can call us; voice messages would also be very convenient.'' [P21]\\
\hline

\addlinespace[0.6em]
\multicolumn{3}{l}{\textbf{Section 5: Additional Financial Behavior}}\\
\hline
Do you know any ways to make money online? &
Tried (1); Scammed (2); Ads on Social Media (Facebook) (1); Don't know (3); Aviator \& S9 (money-making games) (1); Click \& Earn (1); Knows a lot of ways (1); Receives link (1); Click (play) \& earn (7); Earn on TikTok (ads) (3); Knows (3); Smm Lilly - fake likes to make money (1); Scammy apps on TikTok (1); Requires investment (1); No involvement due to not knowing English (1); No (6); Gambling (1); Gets messages (1); TikTok Ads (2); Canva \& Amazon (1) & ``Play games and earn money.'' [P3]\\
\end{longtable}
}

\end{document}